\documentstyle[dina4,11pt]{article}
\newcommand{\be}{\begin{equation}\label}
\newcommand{\ee}{\end{equation}}
\begin{document}
\title{Stationary planar domain walls of a classical spin chain}
\author{ Benno Rumpf 
\\E-mail: benno@arnold.fkp.physik.th-darmstadt.de 
\\Fax: +49 6151 16 4165
\\Institut f\"ur Festk\"orperphysik
\\Technische Hochschule Darmstadt
\\Hochschulstrasse 8
\\64289 Darmstadt
\\Germany}
\maketitle
\begin {abstract}
Domain walls of a discrete model of an anisotropic  ferromagnet are studied. They 
can be described by sequences of two reversible mappings. 
Competition between the length 
scale of spatial structures and the lattice constant
leads to a rich diversity of domain wall solutions related in a
bifurcation scenario.
PACS: 0547,7540C\\
Physics Letters A 221 (1996) 197-203
\end{abstract}

\newpage
\section*{ Introduction}
The complexity of spatial structures of discrete oscillator chains raises 
questions of considerable interest [1]-[8]. 
In this paper domain wall solutions of an one-dimensional 
discrete model of a ferromagnet are contrasted to the well-known Bloch wall 
solution of the 
corresponding continuous system. 
As is known \cite{Bloch}, stationary planar solutions of 
\be{cont1}\dot{\vec{S}}=
\vec{S}\times (J\frac{\partial^2}{{\partial x}^2}\vec{S}+aS_z \vec{e}_z)\ee
are represented by 
the equation $\phi ''= -\frac{a}{J}\cos\phi\sin\phi$ of a point mass in a periodic potential; 
beside spatially periodic solutions, the kink solution 
\be{cont3}\sin \phi = \pm \mbox{tgh} \sqrt{\frac{a}{J}}x\ee 
is obtained. 
However, if the length $\sqrt{\frac{J}{a}}$ of this kink is not large enough, 
the continuum approximation \cite{Kosevich} fails and 
one expects competition between the length scale of a spatial 
structure and the lattice constant. Several papers have studied 
the corresponding classical discrete system \cite{Broek}, \cite{etrich}.\\
This paper is devoted to study of stationary localized structures connecting 
two homogeneous domains in a model of a ferromagnetic chain with one kind of 
atom; quantum corrections \cite{mikeska} are neglected.\\ 
Stationary solutions of oscillator chains with appropriate next neighbour 
coupling can be described by mappings; 
e.g. the Frenkel-Kontorova system is associated to the standard 
map \cite{aubrydaeron},\cite{aubry83}. 
In contrast, complexity of spin chains arises from the fact that the 
coordinates of one spin oscillator are not given unambiguously 
by the position of its two predecessors. For planar solutions two positions of 
the following spin are possible. Consequently only a small part of 
the stationary states can be reached by a single map. At every single step 
one can decide which one of two possible maps is to be applied,  
so that the system may be considered an 
explicitly time-dependent dynamical system. Thus, according to symbolic 
dynamics, a subdivision of the solutions with their defining symbol sequence 
suggests itself. 
Solutions which are characterized by the same inversion symmetry as the Bloch wall are related to symmetry lines of the mappings. \\
\section*{The anisotropic spin chain}
We discuss stationary planar solutions of 
the  model of a spin chain \cite{Lakshmanan},\cite{Belobrov},\cite{Hao} with 
isotropic ferromagnetic coupling and an uniaxial anisotropic term
\be{2b} \dot{\vec{S_n}}=\vec S_n \times (J(\vec S_{n+1}+\vec S_{n-1})+
aS_{n z}\vec e_z) \ee
$J$ is the isotropic Heisenberg coupling constant, $a$ an anisotropy constant;  these constants are real. 
The modulus of the spin is conserved, we set $|\vec{S}_n|=1$. 
$\vec{S}_{n}$ is fixed if the local effective field
$\vec{h}_{eff n}=J(\vec{S}_{n-1}+\vec{S}_{n+1}) +a S_{n z}\vec{e}_z$  is parallel to 
$\vec{S}_{n}$. The position of $\vec S_{n+1}$ can be computed from  
the positions of $\vec{S}_{n}$ and $\vec S_{n-1}$. This determination is not unique; in general several directions $\vec S_{n+1}$ yield parallel 
effective fields $\vec{h}_{eff n}$. In the same way, possible directions 
of $\vec{S}_{n-2}$ can be computed. \\
The system is rotational symmetric with regard to the $z$-axis. Introducing spherical coordinates 
\be{2c} \vec S_n = (\cos \theta_n ,\sin \theta_n \sin \phi_n ,\sin \theta_n \cos  \phi_n)\ee
we investigate solutions with spins perpendicular to the x-axis $(\theta=\frac{\pi}{2})$. 
In this case, fixed point 
conditions are given by the system of equations
\be{2e}\sin (\phi_{n+1}-\phi_{n})=
\sin (\phi_{n}-\phi_{n-1})+\frac{a}{2J}\sin (2\phi_n)\ee
Introducing the second coordinate 
$I_n=\sin (\phi_n-\phi_{n-1})$ 
the fixed point condition can be satisfied equally by
\be{2i}
L_1:  \begin{array}{l}
\phi_{n+1}=\phi_n + \arcsin I_{n+1} \\
I_{n+1}=I_n + k\sin 2\phi_n \\ 
k=\frac{a}{2J}
\end{array}
\ee
and by
\be{2j}
L_2:  \begin{array}{l}
\phi_{n+1}=\phi_n + \pi -\arcsin I_{n+1} \\
I_{n+1}=I_n + k\sin 2\phi_n \\ 
k=\frac{a}{2J}
\end{array}
\ee
Stationary solutions are defined by their initial conditions 
and a sequence of $L_1$'s and $L_2$'s, a property reminiscent 
of symbolic dynamics.  
In general, there are several solutions with the 
same sequence but with different initial conditions. 
In this sense, our system is equivalent to a lowdimensional dynamical 
system with an explicit time dependence which causes the sequence of mappings 
$L_1, L_2$. \\ 
$I_n$ must not exceed 1. For this reason, not all initial  
conditions lead 
to a stationary solution. 
The domain of definition is in general 
a complicated subset of the cylinder $[0,1]\times [0,2\pi]$.\\ 
Each mapping commutes with 
\be{4aaa}
M^-: \begin{array}{l}
\phi '=-\phi \\
I'=-I
\end{array}
\ee
and with 
\be{4aab}
M^{\pi}: \begin{array}{l}
\phi '=\phi +\pi \\
I'=I
\end{array}
\ee

$L_1$ and $L_2$ are reversible mappings \cite{Roberts92}, 
i.e. they can be decomposed into involutions 
$L=H\circ G$ with
$H^2=G^2=Id$ (identity mapping). It follows that 
$L^{-1}=G\circ H$.
Employing $M^-$ and $M^{\pi}$ 
they can each be decomposed into involutions 
in four different ways:
\be{4aac}
L_1=H_1^{0+}\circ G^{0+}=H_1^{\pi +}\circ G^{\pi +}=
H_1^{0-}\circ G^{0-}=H_1^{\pi -}\circ G^{\pi -}
\ee
\be{4aad}
L_2=H_2^{0+}\circ G^{0+}=H_2^{\pi +}\circ G^{\pi +}=
H_2^{0-}\circ G^{0-}=H_2^{\pi -}\circ G^{\pi -}
\ee
We draw up the involutions and symmetry lines $Fix(G), Fix(H)$:
\be{4tab}
\begin{array}{|l|c|} \hline
\mbox{involution} & \mbox{symmetry lines}   \\ \hline
H_1^{0+}: \begin{array}{l}
\phi' =-\phi + \arcsin  I \\
I'=I 
\end{array} 
&  
\phi = \frac{\arcsin  I}{2}
\\ \hline
G^{0+}:
\begin{array}{l}
\phi ' = -\phi \\
I'=I+k\sin 2\phi 
\end{array}
&
\phi =0
\\ \hline
H_1^{\pi +}:
\begin{array}{l}
\phi' = \pi-\phi + \arcsin I \\
I'=I  
\end{array}
&
\phi = \frac{\pi +\arcsin  I}{2}
\\ \hline
G^{\pi +}:
\begin{array}{l}
\phi ' = \pi -\phi \\
I'=I+k\sin 2\phi 
\end{array}
&
\phi = \frac{\pi}{2} 
\\ \hline
H_1^{0-}:
\begin{array}{l}
\phi '=\phi-\arcsin I\\
I'=-I
\end{array}
&
I=0
\\ \hline
G^{0-}:
\begin{array}{l}
\phi '=\phi\\
I'=-I-k\sin 2\phi
\end{array}
&
I=-\frac{k \sin 2\phi}{2}
\\ \hline
H_1^{\pi -}:
\begin{array}{l}
\phi '=\phi-\arcsin I +\pi\\
I'=-I
\end{array}
&
-
\\ \hline
G^{\pi -}:
\begin{array}{l}
\phi '=\phi +\pi\\
I'=-I-k\sin 2\phi
\end{array}
&
-
\\ \hline
H_2^{0+}:
\begin{array}{l}
\phi '=-\phi +\pi -\arcsin I\\
I'=I
\end{array}
&
\phi = \frac{\pi-\arcsin  I}{2} 
\\ \hline
H_2^{0-}:
\begin{array}{l}
\phi '=\phi-\pi +\arcsin I\\
I'=-I
\end{array}
&
-
\\ \hline
H_2^{\pi+}:
\begin{array}{l}
\phi '=-\phi +2\pi -\arcsin I\\
I'=I
\end{array}
&
\phi = \pi -\frac{\arcsin  I}{2} 
\\ \hline
H_2^{\pi-}:
\begin{array}{l}
\phi '=\phi +\arcsin I\\
I'=-I
\end{array}
&
I=0
\\ \hline
\end{array}
\ee
The fact that $L_1$ and $L_2$ have the same involutions $G^{0/\pi\pm}$ 
has a consequence of importance in the further argumentation.  
We assume 
$U$ to be any finite sequence of $L_1$'s and $L_2$'s and $V$ the corresponding 
sequence with reversed order of indices, for instance 
$U=L_2\circ L_2\circ L_1\circ L_2$ 
and $V=L_2\circ L_1\circ L_2\circ L_2$. 
It follows that 
\be{2dd}U^{-1}=G\circ V\circ G\ee
(we skip the indices of $G$ and $H$ if they are unimportant). \\

\section*{Continuous families of domain walls}
In this chapter symmetry properties of domain wall structures 
are studied.  
In our context we understand a domain as a homogeneous or spatially periodic  structure. 
A domain can be defined as a fixed point of a finite sequence 
$D={L_1}^\alpha \circ {L_2}^\beta \circ {L_1}^\gamma ...$ . For example, $(I=0,\phi=0)$ 
is a fixed point of $D=L_1$. This corresponds to the ferromagnetic state 
with parallel spins. $(I=0, \phi=0) $ is also a fixed point of 
$D={L_2}^2$. In this case it corresponds to the antiferromagnetic state of 
antiparallel spins. \\
Domain walls are stationary structures connecting two 
homogeneous domains. 
Let $D_1$, $D_2$ denote two finite sequences of $L_1$ and $L_2$ which 
characterize the domains. 
$(I_{D_1},\phi_{D_1})$, $(I_{D_2},\phi_{D_2})$ are fixed points of  
$D_1$ and $D_2$ respectively. 
A domain boundary solution connecting the two saddle points 
$(I_{D_1},\phi_{D_1})$, $(I_{D_2},\phi_{D_2})$ may contain  
a sequence of mappings different from $D_1$ and $D_2$, so a domain boundary 
solution is defined as a point set determined    
by the sequence $\bar D_1 \circ B \circ \bar D_2$.  
Here $\bar D_1$ and $\bar D_2$ denote infinite sequences of $D_1$ and $D_2$ 
while $B$ is an arbitrary finite sequence of $L_1$ and $L_2$. 
Such a domain boundary solution is obtained if one considers a point sequence emanating from 
a fixed point of $D_1$ on its unstable manifold $W^u_{D_1}$. After the application of a
certain sequence $B$ of the mappings $L_1$ and $L_2$ the point sequence must 
follow again the stable manifold $W^s_{D_2}$ towards a fixed point of $D_2$; 
consequently domain walls are described as heteroclinic orbits of 
the saddle points.\\
In correspondence with the solution of the continuous system, 
we are particularly interested in inversion symmetrical solutions.
Such solutions can only be obtained if $D_1$ and $D_2$ are defined as 
sequences with reversed order of indices, for example 
$D_1=L_1\circ (L_2)^3, D_2=(L_2)^3\circ L_1$ or $D_1=D_2=L_2$.  
$B$ can be an arbitrary finite 
symmetric sequence, for instance 
$(L_2)^2$ or ${L_2}^2\circ L_1\circ {L_2}^2\circ L_1\circ {L_2}^2$. 
$(I_0,\phi_0)$ is a point on the unstable manifold $W^u_{D_1}$ of the 
saddle point $(I_{D_1},\phi_{D_1})= L_1^{-\infty}(I_0,\phi_0)$. 
\be{3baa} S^u_{D_1}(I_0,\phi_0)=\{...(I_0,\phi_0),(I_1,\phi_1),...\}
\subset W^u_{D_1}
\ee 
is the set of all images $D_1^n(I_0,\phi_0), n\in Z$ of $(I_0,\phi_0)$. 
Applying an involution $G$ to this set and using 
$D_2^{-1}=G\circ D_1\circ G$ (\ref{2dd}), $G^2=Id$ 
\be{3bab} S^s_{D_2}=G(S^u_{D_1}(I_0,\phi_0))=
\{...G(I_0,\phi_0),G(I_1,\phi_1),...\}
\ee 
turns out to be the set of iterations $({{D_2}^{-1}})^n(G(I_0,\phi_0))$, 
which is an orbit on the stable manifold $W^s_{D_2}$ of the 
saddle point $G(I_{D_1},\phi_{D_1})$. So the involution $G$ 
transforms the unstable manifold 
of a fixed point into the stable manifold of another fixed point. \\
The sets $S^u_{D_1}$ and $S^s_{D_2}$ represent 
a heteroclinic orbit if they are identical. 
This is true if a common point of 
$S^u_{D_1}$ and $S^s_{D_2}$ exists, e.g. a fixed point of an involution. 
$S^u_{D_1}$ or $S^s_{D_2}$ represent 
an intersection point set of $W^u_{D_1}$ and $W^s_{D_2}$. 
This is the most simple case without a central sequence $B$ different from 
$D_1$ or $D_2$.\\
For a domain wall associated to a sequence $B$, a point 
$(I_B,\phi_B) \in W^u_{D_1}$ must exist which is mapped by $B$ onto $W^s_{D_2}$.
So we must find an intersection point of $B(W^u_{D_1})$ and $W^s_{D_2}$ 
in order to obtain a domain wall solution. 
$B$ must cause  a jump from a point of
$S^u_{D_1}$ to the corresponding point of $S^s_{D_2}$;
 i.e. for one point $(I_B,\phi_B)$ of $W^u_{D_1}$ the conditions
\be{3d}B(I_B,\phi_B)=G(I_B,\phi_B)\ee
must be fulfilled. 
We will show that sequences $\bar{D}_1\circ B\circ \bar{D}_2$ 
lead to structurally stable domain wall solutions.  
There are three different forms of the central symmetric sequence $B$: 
\be{33aaa} 
B=U\circ V 
\ee
\be{33aab}
B=U\circ L_1\circ V
\ee 
\be{33aac}
B=U\circ L_2\circ V
\ee
where $U$ and $V$ have inverse order of indices 
(\ref{2dd}). 
We will show that any intersection point of (\ref{33aaa}) $V(W^u)$ 
or (\ref{33aab}) $L_1\circ V(W^u)$ 
or (\ref{33aac}) $L_2\circ V(W^u)$ and a symmetry line of 
(\ref{33aaa})  $G$, 
(\ref{33aab}) $H_1$ or 
(\ref{33aac}) $H_2$ 
yields an inversion symmetric solution. So to speak, we just have to 
find an intersection point of two lines on a plane; no additional requirement must 
be satisfied for reversibility. \\
For a symmetric solution we have to find a point $(I_B,\phi_B)$ of
$W^u$ with the property (\ref{3d})
We focus the case (\ref{33aaa}) $B=U\circ V$.
It will be shown that condition (\ref{3d}) follows from the existence of 
an intersection point 
$(I_C,\phi_C)=V(I_B,\phi_B)$ of $V(W^u_{D_1})$ and a symmetry line  
of $G^{0/\pi\pm}$. Clearly such a point is invariant under the $G$ concerned.   
With the property (\ref{2dd}) $V^{-1}=G\circ U\circ G$ we have
\be{33c}(I_B,\phi_B)=V^{-1}(I_C,\phi_C)=G\circ U\circ G(I_C,\phi_C)\ee
Applying $G$ and using ${G}^2=Id$, $G(I_C,\phi_C)=(I_C,\phi_C)$ we obtain
\be{33d}G(I_B,\phi_B)=U(I_C,\phi_C)=U\circ V(I_B,\phi_B)=B(I_B,\phi_B)\ee
which had to be proved. \\ 
For the case (\ref{33aab}) $B=U\circ L_1\circ V$ we have to prove 
\be{33e} U\circ L_1\circ V(I_B,\phi_B)=G^{0/\pi\pm}(I_B,\phi_B)\ee
where $(I_B,\phi_B)$ is a point on $W^u_{D_1}$.
Presupposing an intersection point $(I_C,\phi_C)=L_1\circ V(I_B,\phi_B)$ 
of a symmetry line of $H_1^{0/\pi\pm}$ and 
the line $L_1\circ V(W^u)$, we see that 
\be{33f}(I_{C},\phi_{C})=H_1(I_{C},\phi_{C})=
H_1\circ L_1(I_{C-1},\phi_{C-1})=G(I_{C-1},\phi_{C-1})\ee
and consequently 
\be{33g}(I_B,\phi_B)=
V^{-1}(I_{C-1},\phi_{C-1})=G\circ U\circ G(I_{C-1},\phi_{C-1})\ee
Applying $G$ we obtain
\be{33h}G(I_B,\phi_B)=U(I_C,\phi_C)=U\circ L_1\circ V(I_B,\phi_B)\ee
This can be carried out in the same way for (\ref{33aac}) $B=U\circ L_2\circ V$, in which case we presuppose an intersection point of 
a symmetry line of $H_2^{0/\pi\pm}$ and 
the line $L_2\circ V(W^u_{D_1})$. 
We conclude that symmetric solutions constitute a 
continuous family with respect to the parameter $k$.
This proof does not imply that all solutions are symmetric.  
Intersection points of $B(W^u_{D_1})$ and 
$W^s_{D_2}$ may exist which do not correspond to any symmetry line.

\section*{Bifurcations of domain walls}
To put this concept in concrete terms, we focus on solutions which are 
similar to the Bloch wall solution, i.e. inversion 
symmetrical solutions with $D_1=D_2=L_1$ 
belonging to the fixed points $(0,0)$ and $(0,\pi)$ of $L_1$. 
Applying the above considerations we 
identify the solutions by intersection points of $W^u_{D_1}$ or 
$U(W^u_{D_1})$ and symmetry lines.\\
Symmetric domain wall solutions can be divided into two 
classes \cite{etrich}: Configurations with the inversion symmetry point on a 
lattice site are described as central-spin solutions. They correspond 
to central sequences of $B=U\circ V$ (\ref{33aaa}). \\
The remaining (central-bond) configurations are characterized by an inversion 
symmetry point in the middle between two lattice sites. They correspond to 
$B=U\circ L_1\circ V$ (\ref{33aab}) or to 
$B=U\circ L_2\circ V$
(\ref{33aac})\\
In the following, selected solutions described by central sequences $B=L_1$, $L_1^2$ , 
$L_2$ or $L_2^2$ are studied. \\
Typical solutions for moderate parameter values ($k=\frac{a}{2J}=0.28)$ can be inferred from figure~1 which shows the sector $[0,1]\times [0,\pi]$ of the 
phase space. 
The sector can be continued with the period $\pi$ by the symmetry $M^{\pi}$, 
and by $M^-$ it is inversion symmetrical. This corresponds to clockwise and 
anticlockwise solutions. \\
An intersection point of $W^u_{L_1}$ and $W^s_{L_1}$ is mapped to the next 
intersection point but one. Therefore 
two solutions arise from intersection points of $W^u_{L_1}$ and $W^s_{L_1}$. 
Corresponding to (\ref{33d}), 
a heteroclinic set with $B=L_1^2$ exists which contains 
the intersection point $s$ of $W^u_{L_1}$, $W^s_{L_1}$ and the symmetry line 
$\phi= \frac{\pi}{2}$ of $G_1^{\pi +}$. This solution contains a central spin 
with $\phi=\frac{\pi}{2}$ (figure 2).\\
 Corresponding to (\ref{33h}) the heteroclinic set with $B=L_1$ contains an intersection point $b$ of 
$W^u_{L_1}$, $W^s_{L_1}$ and the symmetry line  
$\phi= \frac{\pi + \arcsin I}{2}$ of $H_1^{\pi +}$. 
The center of this solution is in the middle between two lattice sites. 
$(0,0)$ is an intersection point of $W^u_{L_1}$ and 
$Fix(H_2^{\pi +})$; so there exists a sudden kink $\phi_i=0$ for $i<0$, 
$\phi_i=\pi$ for $i>0$. This solution of type $B=L_2$ corresponds 
to the Bloch wall (\ref{cont3}) for $\frac{a}{J}\rightarrow\infty$.\\ 
A differently shaped domain wall of type $B=L_2^2$ corresponds to 
the intersection point $PS1$ of ${L_2}^2 (W^u_{L_1})$ and $W^s_{L_1}$. 
The predecessor $S1$ of $PS1$ 
is an intersection point of 
$L_2(W^u_{L_1})$ and the symmetry line $\phi =\frac{3\pi}{2}$ of $G_1^{\pi +}$, 
so the centre 
of this solution is  $\phi =\frac{3\pi}{2}$ (figure 3).\\
If a localized solution arises from one and the same domain on its right and 
its left side, it is called hump. 
We pick out three simple hump solutions represented by homoclinic orbits 
of the dynamical system. 
The intersection point $C$ of $L_2(W^u_{L_1})$, $W^s_{L_1}$ and the symmetry line 
$\frac{\pi -\arcsin I}{2}$ of $H_2^{0+}$ indicates a homoclinic $B=L_2$ 
solution, which connects $(0,0)$ to itself and is centred at $\phi =\pi$.\\
There exists also a homoclinic solution of $B=L_2^2$ type which can be seen 
from the intersection point $PS2$ of ${L_2}^2 (W^u_{L_1})$ and $W^s_{L_1}$. 
The predecessor $S2$ 
of $PS2$ is an intersection point of $L_2(W^u)$ and the symmetry line 
$\phi=\pi$ of $G^{0+}$; the 
solution is centred at $\phi=\pi$. The origin $(0,0)$ is also an intersection 
point of these two lines which corresponds to a trivial hump 
$\phi_i=0,\phi_0=\pi,\phi_j=0 ,i<0, j>0$
where only one spin is folded down in relation to the ferromagnetic state.\\
In the following, bifurcations of these solutions will be briefly 
described. To determine the stability of solutions in question the equations 
of motion for a number (usually 30-50) of spins of the domain wall
are linearized while the boundaries are 
fixed. Eigenvalues of this matrix are computed numerically. Due to the 
rotation symmetry of the problem, two eigenvalues are zero. The solution 
corresponds to a local energy minimum if all other eigenvalues are imaginary. This fixed point becomes attractive if a phenomenological Landau-Lifshitz damping term \cite{Rumpf} is added.\\ 
The solution corresponding to intersection point $b$ is stable. It exists up 
to $k=0.53$, where the maximal $I$ of the $L_1$ 
solution (point $b$) reaches 1. $L_1$ and $L_2$ have the same effect at this point.  It follows that a $B=L_1$ solution is 
identical with a $B=L_2$ solution. 
Indeed, with this parameter value a branch of $L_2(W^u_{L_1})$ touches 
$W^s_{L_1}$. 
So the $B=L_1$ solution merges into a 
$B=L_2$ solution with an opening angle larger than $\frac{\pi}{2}$. 
This solution is stable as well.\\
For increasing $k$, the intersection point of 
$L_2 (W^u_{L_1})$ and $W^s_{L_1}$ is moving 
down to $I=0,\phi=\pi$. 
At this point, the clockwise $L_2$ 
solution and its anticlockwise counterpart merge into the sudden kink solution 
in a pitchfork-bifurcation. In this bifurcation the sudden kink solution 
corresponding to the intersection point $(0,0)$ of $W^u_{L_1}$ and 
$Fix(H_2^{\pi +})$ becomes stable. 
We can compute the parameter value of this bifurcation analytically. 
The slopes of $W^s_{L_1}$ and of $L_2 (W^u_{L_1})$ at $(0,\pi)$ must coincide, i.e.
\be{4g} 
-k-\sqrt{2k+k^2}=
\frac{k+\sqrt{2k+k^2}}
{1-k-\sqrt{2k+k^2}}
\ee
must be satisfied, which happens at $k=\frac{2}{3}$. This is the analytical 
confirmation of the numerically determined value in \cite{Broek}.\\
The solution containing a central spin $s$ is stable and has a slightly 
higher energy. It exists for all 
parameter values. For 
$\frac{a}{J}\rightarrow\infty$ we get $\phi_i\rightarrow 0$ for $i<0$, 
$\phi_i=\frac{\pi}{2}$ for $i=0$ $\phi_i\rightarrow\pi$ for $i>0$.\\
The solution corresponding to the intersection point $S2$ appears in a 
pitchfork-bifurcation out of a trivial solution  
$\phi_i=0,\phi_0=\pi,\phi_j=0 ,i<0, j>0$. 
At the critical parameter value the gradients 
of $W^s_{L_1}$ and ${L_2}^2 (W^u_{L_1})$ must coincide.  
The value is computed analogously to (\ref{4g}), we get $k=\frac{1}{4}$.
For $k<1.28$ the trivial hump is unstable, for larger $k$ it is stable. 
All other solutions of figure 3 are unstable.



\section*{Conclusion}

In contrast to the related continuous system, the discrete spin chain has a large variety of stationary planar solutions. As a peculiarity of spin systems, 
a virtual sequence of two mappings generates a stationary solution. Consequently a
solution is defined both by its initial conditions and a generating symbol 
sequence. 
Besides a variety of solutions with no correlate in the continuous model,  the discrete spin chain has several solutions similar to the Bloch wall solution. The latter leads to degeneration if the continuum 
limit is a good approximation (weak anisotropy): Two stable solutions of the discrete system converge 
to the same continuum limit. On the other hand, for strong anisotropy, 
one of these solutions bifurcates into an abrupt kink of 180 degrees. 
Among the immense number of solutions of a nonlinear mapping, chaotic 
orbits can be identified with stationary states of 
the spin chain. For moderate values of $k$, some KAM-tori have 
already vanished but others still ensure that 
the trajectory does not leave the definition area. Because of ener\-getical 
instability, the physical relevance of these solutions should be assessed with care. Even stable solutions may be unobservable due to their high energy, 
but also unstable stationary solutions may be relevant in dynamics. 
Firstly, unstable stationary and periodic solutions contribute to the structure of chaotic attractors and are relevant for their construction \cite{Ruelle}. Secondly it was suggested \cite{Eckmann} that unstable stationary solutions 
influence transient dynamics in the process of pattern formation.

\section*{Acknowledgement}
The author wishes to thank  Professor H.Sauermann for valuable  
discussions with him. He is also indebted to 
Dr.W.Just for his critical 
reading of the manuscript and his clarifying comments.
This work was performed within a program of Sonderforschungsbereich 185 
Nichtlineare Dynamik, Frankfurt, Darmstadt. 

\newpage
\section* {Figure captions}
Figure 1\\
The sector $[0,1]\times [0,\pi]$ of the phase space for $k=0.28$. It shows $W^u$, $W^s$, $L_2(W^u)$, ${L_2}^2(W^s)$ 
and all symmetry lines. 
Here $L_2(W^u)$ is the image of $W^u$ of the segment 
$\phi\in [-\pi , 0]$. 
There are a central-bond set of intersection points 
of $W^u$ and $W^s$ including the intersection point $b$ of $W^u$, $W^s$, 
$Fix(H_2^{\pi +})$ and a central-spin set including the intersection point $s$ of $W^u$, $W^s$, $Fix(G^{\pi +})$.  
The intersection point $B$ of $L_2(W^u)$, $W^s$ and $Fix(H_2^{0+})$ 
belongs to a central-bond solution. 
There are two other central-spin solutions: One of them corresponds to 
the intersection point $PS1$, its predecessor $S1$ is a intersection point of 
$L_2(W^u)$ and the symmetry line $\phi=\pi$. The predecessor $S2$ of the  
point $PS2$ is an intersection point of $L_2(W^u)$ and $\phi =0$. 
The origin of this solution is a pitchfork bifurcation at $k=0.25$ where the gradients of
$W^s$ and ${L_2}^2(W^u)$ at $(0,\pi)$ coincide.\\\\
Figure 2\\
The angle $\phi$ of three basic domain wall solutions as a function of 
the spin index N at $k=0.28$.  
The kinks corresponding to symmetry lines $G^{\pi +}$ and $H^{\pi +}_{1}$ 
are energetically stable, these solutions are generated only by $L_1$. 
The sudden kink corresponding to the symmetry line $G^{0+}$ is caused by 
a central application of $B=L_2$. It becomes stable at $k=\frac{2}{3}$ 
(weak coupling and virtually isolated spins). \\\\
Figure 3\\
The domains $\phi=0$ and $\phi=\pi$ are linked by a solution corresponding 
to the symmetry line $G^{\pi +}$ (note $\phi$ mod$2\pi$). 
Discernible from opening angles larger than $\frac{\pi}{2}$, 
this solution contains a central sequence $B={L_2}^2$. So $PS1$ is generated 
from its predecessor $S1$ by $L_2$. Similarly, the hump $G^{0+}$ contains 
$B={L_2}^2$, while the hump $H^{0+}_2$ contains $B=L_2$. At the trivial hump 
$G^{0+}$, only one spin $S2$ is folded down in relation to the ferromagnetic domain. This 
solution becomes stable for weak coupling $(k=1.28)$ while all other solutions 
in this picture are unstable.

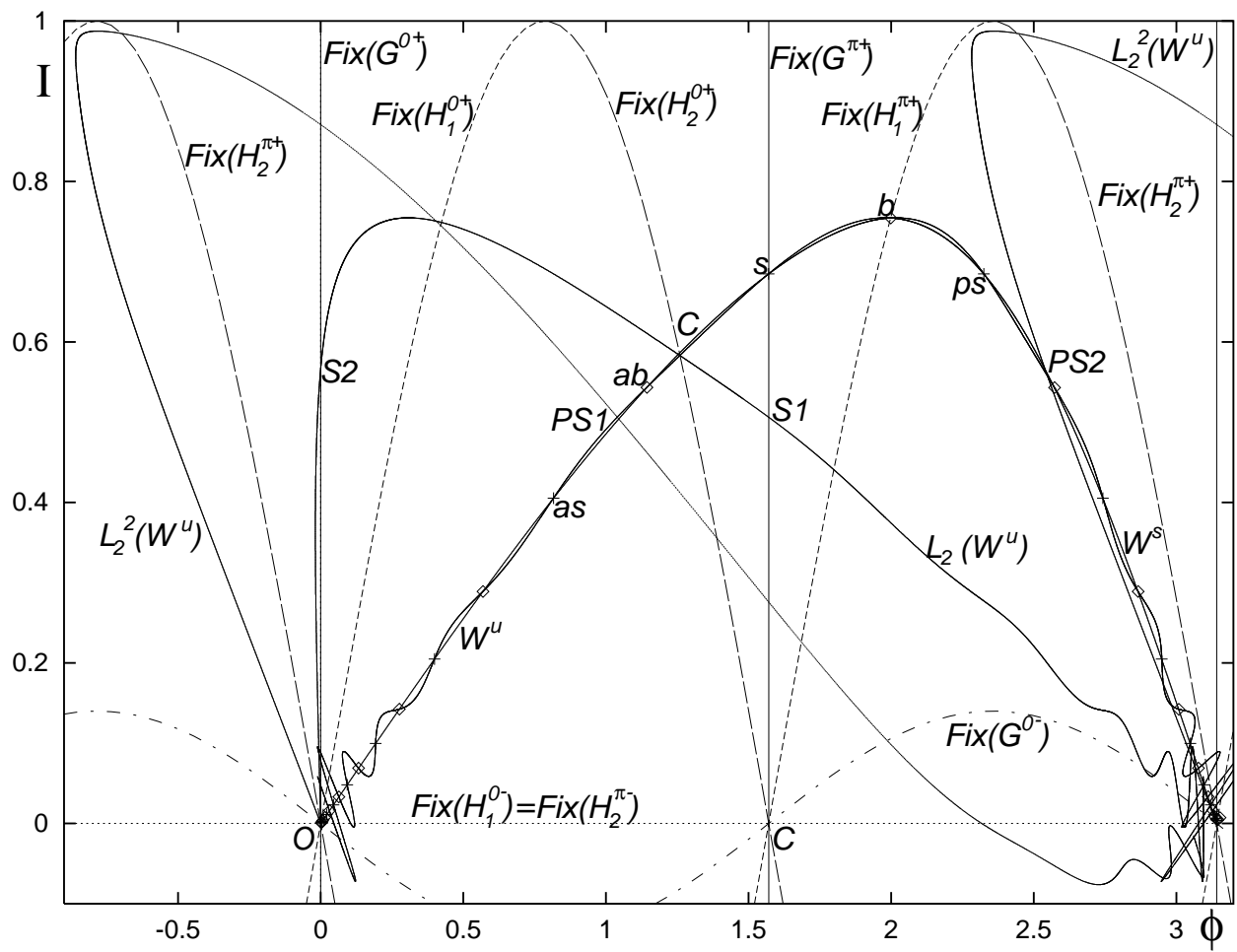
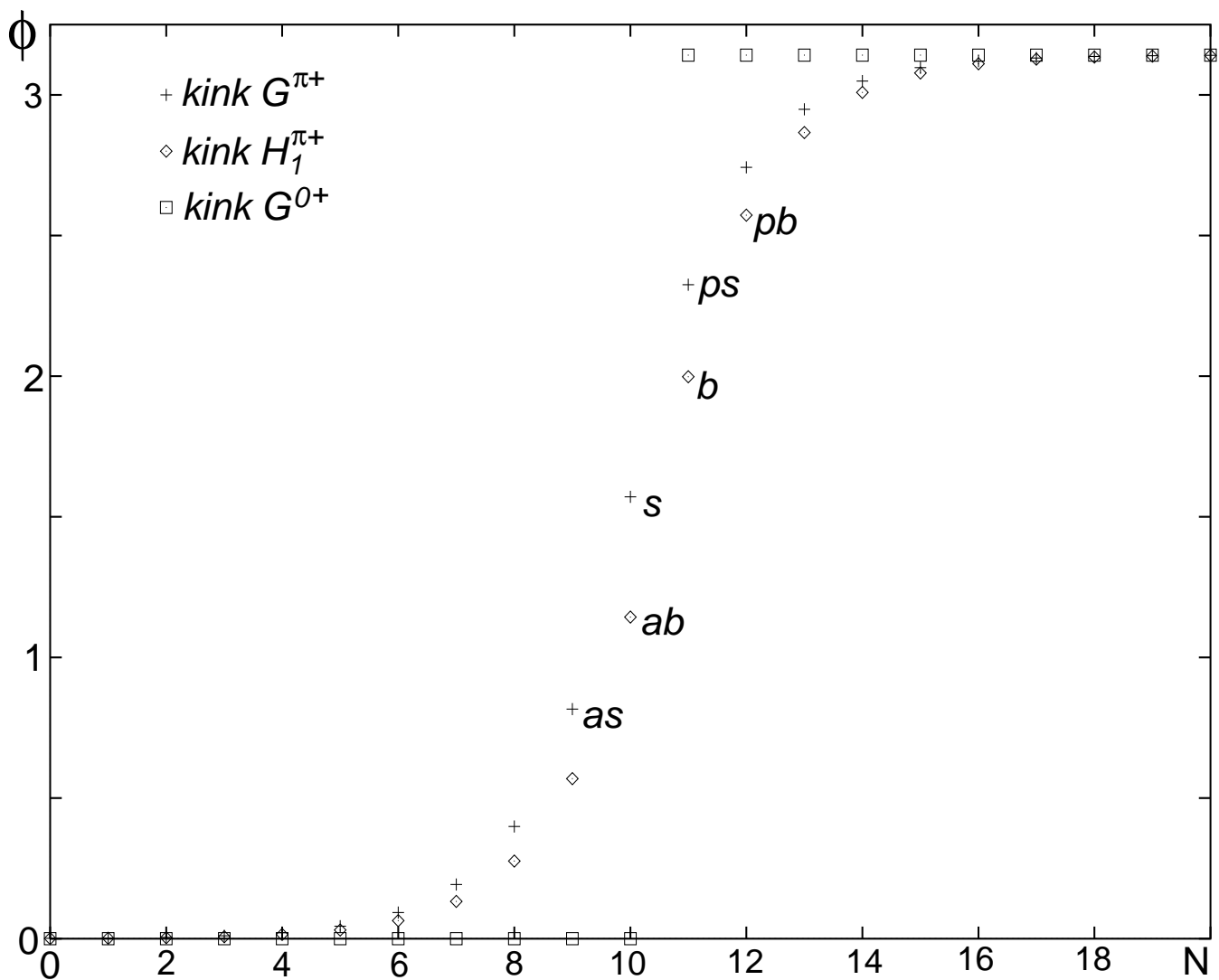
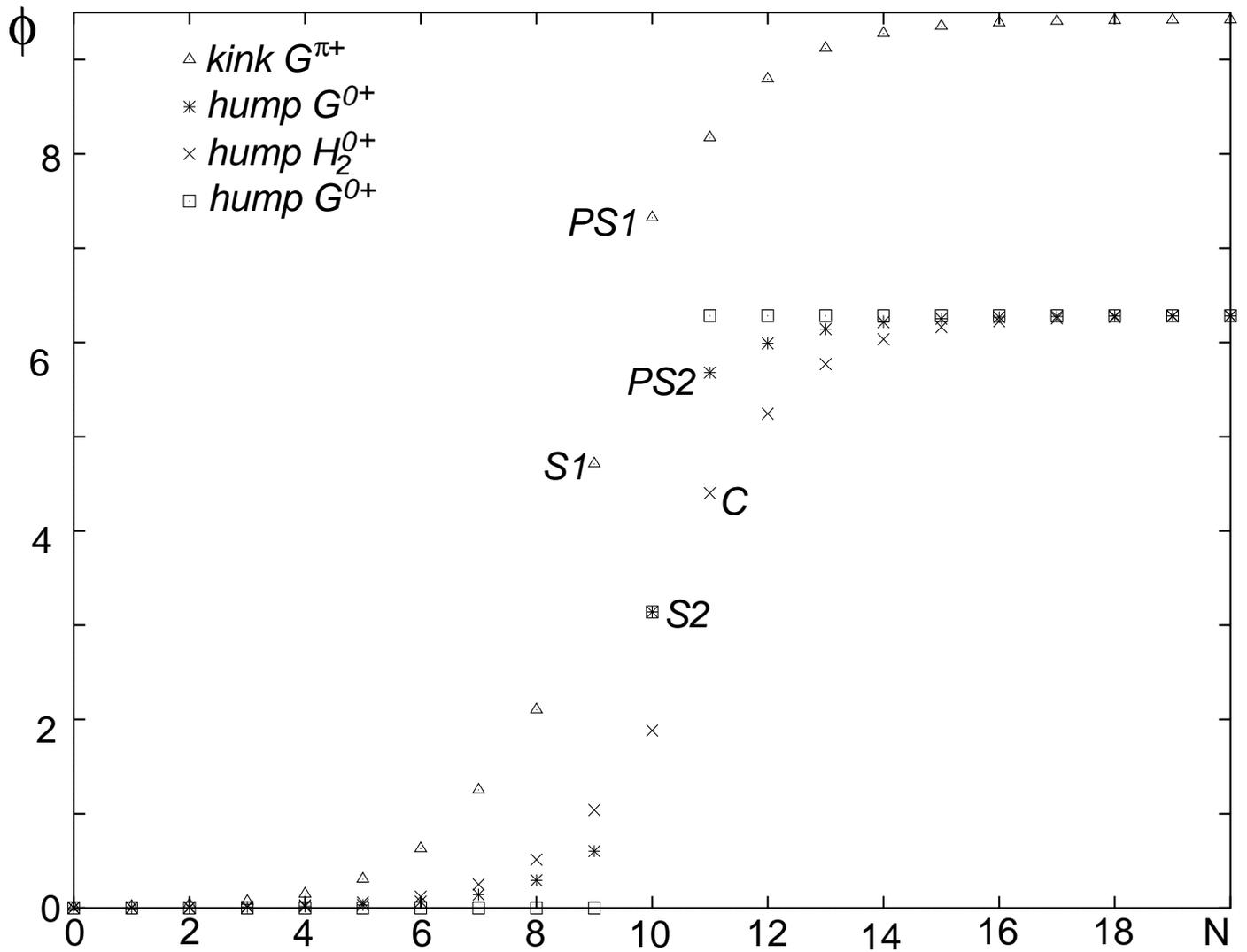
\end{document}